\def\rfr#1{eq. (\ref{#1})}

\def\virg#1{``#1''}

\def\bb#1#2#3{\bibitem[\protect\citeauthoryear{#1}{#2}]{#3}}
\def\eqi{\begin{equation}}
\def\eqf{\end{equation}}
\def\eqia{\begin{eqnarray}}
\def\eqfa{\end{eqnarray}}

\def\rp#1#2{{#1\over#2}} \def\lb#1{\label{#1}}

\def\bds#1{\boldsymbol{#1}}

\documentclass{elsart}

\usepackage{slashbox}
\usepackage{url}
\usepackage{amsmath}
\usepackage{latexsym,wasysym}

\usepackage{natbib}

\usepackage{epsfig}

\usepackage{amssymb}

\RequirePackage{color}

\begin{document}

\begin{frontmatter}



\title{Dynamical orbital effects of General Relativity on the satellite-to-satellite range and range-rate in the GRACE mission: a sensitivity analysis
}


\author{Lorenzo Iorio\thanksref{footnote2}}
\address{Ministero dell'Istruzione, dell'Universit\`{a} e della Ricerca (M.I.U.R.)-Istruzione\\
International Institute for Theoretical Physics and
High Mathematics Einstein-Galilei\\ Fellow of the Royal Astronomical Society (F.R.A.S.)}
\thanks[footnote2]{Address for correspondence:  Viale Unit\`{a} di Italia 68, 70125, Bari (BA), Italy}
\ead{lorenzo.iorio@libero.it}
\ead[url]{http://digilander.libero.it/lorri/homepage$\_$of$\_$lorenzo$\_$iorio.htm}



\begin{abstract}
We numerically investigate the  impact of the General Theory of Relativity (GTR) on the orbital part of the satellite-to-satellite range $\rho$ and range-rate $\dot\rho$ of the twin GRACE A/B spacecrafts through their \textcolor{black}{post-Newtonian (PN)} dynamical equations of motion integrated in an Earth-centered frame over a time span $\Delta P=1$ d.
%
%
%
%
%
The present-day accuracies in measuring the GRACE biased range and range-rate are $\sigma_{\rho}\sim 1-10\ \mu$m, $\sigma_{\dot\rho}\sim 0.1-1\ \mu$m s$^{-1}$.
%
%
%
%
%
%
The GTR range and range-rate effects turn out to be $\Delta\rho=80$ $\mu$m and $\Delta\dot\rho=0.012$ $\mu$m s$^{-1}$ (\textcolor{black}{1PN gravitomagnetic}), and $\Delta\rho=6000$ $\mu$m and $\Delta\dot\rho=10$ $\mu$m s$^{-1}$ (\textcolor{black}{1PN gravitoelectric}).
\textcolor{black}{It turns out that the range shifts $\Delta \rho$ corresponding to the GTR-induced time delays $\Delta t$ on the propagation of the electromagnetic waves linking the GRACE spacecrafts are either negligible (1PN gravitomagnetic) or smaller (1PN gravitoelectric) than the orbital effects by about 1 order of magnitude over $\Delta P=1$ d.}
 We also compute the dynamical range and range-rate perturbations caused by the first six  zonal harmonic coefficients $J_{\ell}, \ell=2,3,4,5,6,7$ of the classical multipolar expansion of the \textcolor{black}{geopotential}  to evaluate their aliasing impact on the relativistic effects. Conversely, we quantitatively, and preliminarily, assess the possible a-priori \virg{imprinting} of GTR itself, not solved-for in all the GRACE-based Earth's gravity models produced so far, on the low degree zonals of the geopotential. The present sensitivity analysis  can also be extended, in principle, to different orbital configurations in order to design a suitable dedicated mission able to accurately measure GTR. Moreover, it may be the starting point for more refined numerical investigations concerning the actual measurability of the relativistic effects involving, e.g., a simulation of full GRACE data, including GTR \textcolor{black}{itself}, and the consequent parameters' estimation. Finally, also other non-classical dynamical features of motion, caused by, e.g., modified models of gravity, may be considered in further studies.
\end{abstract}


\end{frontmatter}

\parindent=0.5 cm

\section{Introduction}
The Gravity Recovery and Climate Experiment (GRACE) mission \citep{grace0,grace1,grace2}, jointly launched in March 2002 by NASA and the German Space Agency (DLR) to  map the terrestrial gravitational field with an unprecedented accuracy, consists of a\footnote{\textcolor{black}{The idea of using the intersatellite signal between a pair of Earth orbiting spacecrafts to accurately measure certain features of the terrestrial gravitational field  dates back to \citet{wolf}. The first mission concepts were proposed by \citet{early1} (GRAVSAT), and  \citet{early2} (SLALOM).}} tandem of two spacecrafts moving along low-altitude, nearly polar orbits (see
Table \ref{stateveckep} \textcolor{black}{for their orbital parameters}) continuously linked by an inter-satellite microwave K-band ranging (KBR) system accurate to better than $10\ \mu$m (biased range $\rho$) \citep{range} and $1\ \mu$m s$^{-1}$ (range-rate) \citep{range,rrate}. Investigations concerning a follow-on of the GRACE mission are being currently performed \citep{follow2}; by using an interferometric laser ranging system it would be possible to reach an  accuracy level of $\sim$nm s$^{-1}$ or better in measuring the range-rate \citep{follow1}.

Although GRACE was not specifically designed to directly test the General Theory of Relativity (GTR), which, \textcolor{black}{indeed}, was never solved-for in the several global gravity field solutions\footnote{See http://icgem.gfz-potsdam.de/ICGEM/ on the WEB.} produced so far by different institutions from long data records from GRACE, the great accuracy in its KBR may, in fact, allow, in principle, to measure some consequences of GTR  by exploiting such direct accurate observables. Concerning the Lense-Thirring effect \citep{LT}, connected with the rotation of the source of the gravitational field \textcolor{black}{(see \rfr{gemeq} and \rfr{egos} below)}, a similar idea was envisaged by \citet{mash}. Thus, it is important to investigate the impact of the dynamical effects of GTR on both range and range-rate to preliminarily check if it falls within the present-or future-sensitivity domain of GRACE-type missions. It will be the subject of Section \ref{seconda} \textcolor{black}{in which both numerical and analytical calculations will be performed}. \textcolor{black}{To this aim, we stress that we will deal in depth with the effects of GTR on the range and range-rate of GRACE coming only from the orbital motions of the twin spacecrafts. \textcolor{black}{Indeed, as shown in Section \ref{delay}, the range perturbations corresponding to the GTR time delays affecting the propagation of the electromagnetic waves mutually linking GRACE A and GRACE B are not particularly important for our purposes. About the Lense-Thirring time delay \citep{Kope,kope97,kope02,ciufo03}, in Section \ref{delayLT} we will show that it is completely negligible for GRACE. The Shapiro time delay \citep{shapiro64} is, instead, large enough to be detectable, but the magnitude of the corresponding range shift is smaller by about one order of magnitude than that due to the orbital motions (Section \ref{delayGE})}. It should also be noticed that we do not aim to quantitatively
assess the actual measurability of the \textcolor{black}{post}-Newtonian dynamical effects investigated. It is a different
and important task which would deserve a dedicated work. Indeed, a fit of the
initial conditions with quite  a lot of other dynamical and empirical parameters to the real (or realistically simulated) observations would be needed in order to plausibly
evaluate the level of removal of the effects of interest from the signatures. It is a
nontrivial task which is beyond the scope of the present analysis.}
Concerning other, clock-related relativistic effects in GRACE, it must be recalled that dual-frequency carrier-phase Global Positioning System (GPS) receivers are flying on both satellites. They are used  for precise orbit determination of both the GRACE A/B spacecrafts, and to time-tag the KBR system;
the relativistic effects in the GRACE GPS data were examined by \citet{ash}.

Some of the Earth's gravity field solutions retrieved from GRACE data were used as background reference models in the LAGEOS-based tests \citep{ciuf,ior} of the Lense-Thirring effect \citep{LT}; the even ($\ell=2k, k=1,2,3,...$) zonal ($m=0$) harmonic coefficients $J_{\ell}$ of the multipolar expansion of the classical part of the terrestrial gravitational potential, estimated as solve-for parameters in the GRACE-based models, may retain an a-priori \virg{imprinting} by GTR itself which, as already noted, has never been explicitly solved-for so far in the GRACE data processing. This aspect will be treated in Section \ref{terza}. Section \ref{conclu} is devoted to the conclusions.

Finally, let us remark that the approach followed here in the specific case of GTR can well be extended to other dynamical effects predicted, e.g., by modified models of gravity. A first example is the investigation of the effects of a Yukawa-like extra-force on the orbit of GRACE-A   performed by \citet{apss}.
\textcolor{black}{Such a perspective has recently gained a new appeal after the proposal by \citet{dvali} about a putative Earth-sourced fifth force as possible explanation of  observed  neutrinic phenomenology in some Earth-based laboratories \citep{opera}.}
\textcolor{black}{\section{General relativistic effects in the satellite-to-satellite range and range-rate}\lb{seconda}
The relativistic treatment of the near-Earth satellite orbits, as far as the equations of motion are concerned, is outlined in \citet[p. 155]{IERS} and references therein.}
\textcolor{black}{\subsection{Numerical calculation}}
The \textcolor{black}{numerical} approach adopted is as follows.

We simultaneously integrated the equations of motion of both the GRACE spacecrafts in a geocentric local inertial frame\footnote{We neglected non-inertial effects caused in it by tidal forces due to external bodies of the solar system \citep{KopeBrum}.} endowed with cartesian coordinates by including the GTR effects we are interested in as dynamical  orbital perturbations of the Newtonian monopole.
\textcolor{black}{More specifically,
the Parameterized Post-Newtonian (PPN) perturbing acceleration $\bds A_{\rm PPN}$ to be added to the Newtonian monopole $\bds A_{\rm Newton}\doteq -GM \bds{\hat{r}}/r^2$ in the equations of motion
\eqi \rp{d^2 \bds r}{dt^2}= \bds A_{\rm Newton}+\bds A_{\rm PPN}\eqf
is\footnote{\textcolor{black}{It is assumed that the de Sitter precession has been transformed away.}}, to order $\mathcal{O}(c^{-2})$ (1PN), \citep[p. 89, p.95]{Soffel}, \citep[p. 155]{IERS}
\eqi \bds A_{\rm PPN}\doteq-\bds E_g -2\left(\rp{\bds v}{c}\right)\times\bds B_g.\lb{gemeq}\eqf
In it \citep[p. 89, p.95]{Soffel}, \citep[p. 106]{IERS},
\eqi
\left\{
\begin{array}{lll}
\bds E_g &\doteq & -\rp{GM}{c^2 r^3}\left\{\left[\rp{2(\beta+\gamma)GM}{r}-\gamma v^2\right]\bds r+2(1+\gamma)\left(\bds r\cdot\bds v\right)\bds v\right]\}, \\ \\
\bds B_g &\doteq & -\left(\rp{1+\gamma}{2}\right)\rp{GS}{cr^3}\left[\bds{\hat{S}} - 3\left(\bds{\hat{S}}\cdot{\bds{\hat{r}}} \right){\bds{\hat{r}}}\right],\lb{egos}
\end{array}
\right.
\eqf
where $\gamma,\beta$ are the usual PPN parameters \citep{Will},   $c$ denotes the speed of light in vacuum, $G$ is the Newtonian constant of gravitation, $M$ and $\bds S$ are the mass and the proper angular momentum, respectively of the central body, and $\bds v$ is the velocity of the test particle moving at distance $r$ from $M$; the unit vector $\bds{\hat{r}}$ is directed from the central body to the test particle.
In \rfr{gemeq}-\rfr{egos} $\bds E_g$ is the so-called \virg{gravitoelectric}, Schwarzschild-like field, while  $\bds B_g$ is the \virg{gravitomagnetic} one yielding, among other things, the \citet{LT} orbital precessions. We worked in the GTR case, i.e. for $\gamma=\beta=1$.}

Then, we performed another integration without them.

The time span of both the integrations was $\Delta P=1$ d\textcolor{black}{, in order to keep contact with the actual data reduction procedures followed by the GRACE analysts \citep{Flech011}}. The method adopted, implemented with the software package MATHEMATICA, is the \textcolor{black}{ExplicitRungeKutta one, with ${\rm MaxSteps}\rightarrow 10^6$ and ${\rm MaxStepFraction}\rightarrow1/1000$.}
The initial conditions, common to all the numerical integrations, are in Table \ref{stateveckep}.
\begin{table*}[ht!]
\caption{Keplerian orbital elements of the pair GRACE A/B corresponding to \textcolor{black}{the state vectors, in cartesian coordinates, of the files GNV1B$\_$2003-09-14$\_$A$\_$00 and  GNV1B$\_$2003-09-14$\_$B$\_$00 retrieved from ftp://cddis.gsfc.nasa.gov/pub/slr/predicts/current/graceA$\_$irvs$\_$081202$\_$0.gfz and ftp://cddis.gsfc.nasa.gov/pub/slr/predicts/current/graceB$\_$irvs$\_$081201$\_$1.gfz. The epoch is 13 September 2003. See ftp://podaac.jpl.nasa.gov/pub/grace/doc/Handbook$\_$1B$\_$v1.3.pdf for the explanation of the GPS Navigation Data Format Record (GNV1B) format. The orbital elements} are the semi-major axis $a$, the eccentricity $e$, the orbital inclination $I$ to the Earth's equator, the longitude of the ascending node $\Omega$, the argument of pericenter $\omega$, and the mean anomaly $\mathcal{M}$. The Keplerian orbital periods $P^{\rm (Kep)}\doteq 2\pi\sqrt{a^3/GM_{\oplus}}$ of the GRACE pair are of the order of $\approx 1.56$ h$= 0.065$ d.
}\label{stateveckep}
\centering
\bigskip
\begin{tabular}{lllllll}
\hline\noalign{\smallskip}
S/C & $a_0$ (km) & $e_0$  & $I_0$ (deg) & $\Omega_0$ (deg) & $\omega_0$ (deg) & $\mathcal{M}_0$ (deg)\\
\noalign{\smallskip}\hline\noalign{\smallskip}
A & $6841.11877$ & $0.00272831$ & $89.9395$ & $-71.5742$ & $119.916$ & $-179.997$ \\
B &  $6839.80210$ & $0.00298412$ & $89.8374$ & $-71.5081$ & $118.082$ & $-179.997$\\
\noalign{\smallskip}\hline\noalign{\smallskip}
\end{tabular}
\end{table*}
The altitudes of the twin GRACE spacecrafts are  about $500$ km with respect to the Earth's surface; their orbits are almost circular and polar.

The resulting numerically integrated trajectories were, then,  used to compute the satellite-to-satellite range perturbation $\Delta\rho$ as the difference among the perturbed and the unperturbed ranges. The range-rate perturbation $\Delta\dot\rho$ was straightforwardly computed by numerically differentiating $\Delta\rho$.

\subsection{Analytical calculation}\lb{anale}
Our numerical approach was successfully tested by comparing its outcome for the Lense-Thirring effect to an analogous analytical calculation of the gravitomagentic range and range-rate signals. Here we outline it in detail.

\textcolor{black}{
 As a first step, it is required to work out the short-period, i.e. not averaged over one full orbital revolution, Lense-Thirring  perturbations of all the six osculating Keplerian orbital elements $a,e,I,\Omega,\omega,{\mathcal{M}}$ of a test particle. It can  be done in the framework of the standard\footnote{\textcolor{black}{Here $R,T,N$ are the radial, transverse and out-of-plane (denoted also as normal) directions of the orthonormal frame co-moving with the test particle. Such a frame is usually adopted to decompose any perturbing acceleration acting on the particle itself. See \rfr{krizza}-\rfr{krozza} below.}} $R-T-N$ formalism \citep[p. 90]{Soffel} applied to the perturbative Gauss equations \citep[p. 25]{Joos}, \citep[p. 90]{Soffel} for the variation of the orbital elements. By using the $R-T-N$ components  of the gravitomagnetic force \citep[p. 95]{Soffel}, \citep[p. 24]{Joos}  in \rfr{gemeq}, computed in a frame with the $z$ axis directed along $\bds{S}$, it is possible to obtain\footnote{\textcolor{black}{Similar results can be found in \citet[p. 95]{Soffel}, but they refer to the case  $f_0=0$.}  }
\begin{equation}
\left\{
\begin{array}{lll}
\Delta a_{\rm LT} & = & 0, \\ \\
\Delta e_{\rm LT} & = & -\rp{2GS\cos I\left(\cos f - \cos{f_0}\right)}{c^2 n a^3 \sqrt{1-e^2}},\\  \\
\Delta I_{\rm LT} & = & -\rp{2GS \sin I \left[\left(1+e\cos f\right)\cos^2 u-\left(1+e\cos{f_0}\right)\cos^2{u_0}\right] }{c^2 n a^3 (1-e^2)^{3/2}}, \\  \\
\Delta \Omega_{\rm LT} & = &  \rp{ GS \left\{2\left[\left(f-f_0\right) +e\left(\sin f - \sin f_0\right) \right] -\left[\left(1+e\cos f\right)\sin 2 u-\left(1+e\cos f_0\right)\sin 2 u_0\right] \right\} }{ c^2 n a^3  (1-e^2)^{3/2} },\\ \\
\cos I\Delta \Omega_{\rm LT} +\Delta\omega_{\rm LT} & = &-\rp{2GS\cos I \left[2e \left(f-f_0\right) + (1+e^2)\left(\sin f-\sin f_0\right)\right]}{c^2 n e a^3 (1-e^2)^{3/2}}, \\  \\
\Delta {\mathcal{M}}_{\rm LT} & = & \rp{2GS\cos I \left(\sin f-\sin {f_0}\right)}{c^2 n e a^3 }.
\end{array}\lb{equazz}
\right.
\end{equation}
 In \rfr{equazz} $n\doteq\sqrt{GM/a^3}$ is the unperturbed Keplerian mean motion, $f$ is the true anomaly reckoning the instantaneous position of the test particle along the Keplerian ellipse, and $u\doteq \omega + f$ is the argument of the latitude;
 it is intended that $u_0\doteq f_0+\omega$ since $\omega$ is fixed for an unperturbed  Keplerian orbit, where $f_0$ is the true anomaly at the epoch.
Then, by means of the general relations \citep{Caso},
\begin{equation}
\left\{\begin{array}{lll}
\Delta R &=&\left(\rp{r}{a}\right)\Delta a - a \cos f\Delta e +\rp{ae\sin f}{\sqrt{1-e^2}}\Delta{\mathcal{M}}, \\ \\
\Delta T &=& a\sin f\left[1+\rp{r}{a(1-e^2)}\right]\Delta e + r(\cos I\Delta\Omega + \Delta\omega)+\left(\rp{a^2}{r}\right)\sqrt{1-e^2}\Delta{\mathcal{M}},\\ \\
\Delta N &=& r(\sin u\Delta I-\cos u\sin I\Delta\Omega),
\end{array}\lb{dnorm}
\right.
\end{equation}
it is possible to analytically work out the Lense-Thirring perturbations of the $R-T-N$ components of the orbit. They turn out to be
\eqi
\left\{
\begin{array}{lll}
  \Delta R_{\rm LT} &=& \rp{2GS\cos I\left[1-\cos\left(f-f_0\right)\right]}{c^2 n a^2 \sqrt{1-e^2}},\\  \\
  \Delta T_{\rm LT} &=&  -\rp{2GS\cos I \left\{2\left[\left(f-f_0\right) -\sin\left(f-f_0\right)\right]  +e\left[1-\cos\left(f-f_0\right)\right]\sin f  \right\}  }{c^2 n a^2 \sqrt{1-e^2}(1+e\cos f)},\\  \\
  \Delta N_{\rm LT} &=& \rp{2GS\sin I  \left\{  \left(1+e\cos f_0\right)\cos u_0 \sin\left(f-f_0\right) -\left[\left(f-f_0\right) +e\left(\sin f -\sin f_0\right)\right]\cos u \right\} }{c^2 n a^2 \sqrt{1-e^2}(1+e\cos f)}.
\end{array}\lb{inculescion}
\right.
\eqf
Note that the result of \rfr{inculescion}, which is novel, is also an exact one; no approximations in $e$ were used. Moreover, \rfr{inculescion} does not present any singularities for particular values of  $e$ and  $I$.
An inspection of \rfr{inculescion} shows that the radial perturbation $\Delta R_{\rm LT}$ consists of the sum of a constant offset and a 1-cycle-per-revolution (cpr) harmonic term; no cumulative, secular components are present, so that the average radial shift is simply given by the constant term.
Instead, in the transverse shift $\Delta T_{\rm LT}$ a dominant secular term is present in addition to a 1-cpr harmonic one; both are of zero order in $e$. There is also a smaller, harmonic component of order $\mathcal{O}(e)$. The normal perturbation $\Delta N_{\rm LT}$ has, at zero order in $e$,  a secular term, whose amplitude is modulated by $\cos u$, and a 1-cpr harmonic component; a smaller, harmonic term of order   $\mathcal{O}(e)$ is present as well.
While $\Delta R_{\rm LT}$ and $\Delta T_{\rm LT}$ vanish for polar orbits, i.e. for $I=90$ deg, $\Delta N_{\rm LT}$ is zero for equatorial orbits, i.e. for $I=0$ deg.
}

\textcolor{black}{
In order to conveniently plot \rfr{inculescion} as a function of time, we will use the useful relation for the guiding center \citep[p. 41]{Murr}
\eqi
\begin{array}{c}
f\simeq {\mathcal{M}} + 2e\sin {\mathcal{M}} +\rp{5}{4}e^2\sin 2{\mathcal{M}}+\\  \\
+ e^3\left(\rp{13}{12}\sin 3{\mathcal{M}} -\rp{1}{4}\sin {\mathcal{M}}\right)+e^4\left(\rp{103}{96}\sin 4{\mathcal{M}} - \rp{11}{24}\sin 2{\mathcal{M}}\right);
\end{array}\lb{anomazza}
\eqf
indeed, ${\mathcal{M}}\doteq n(t-t_p),$ where $t_p$ is the time of the passage at pericenter.}

\textcolor{black}{
The results of \rfr{inculescion}, with \rfr{anomazza}, are the basic elements to work out the Lense-Thirring perturbation $\Delta\rho_{\rm LT}$ of the two-body range $\rho$. Indeed, from \citep{cheng}
\eqi
\left\{
\begin{array}{lll}
\rho^2 &=&\left(\bds{r}_{\rm A}-\bds{r}_{\rm B}\right)\cdot\left(\bds{r}_{\rm A}-\bds{r}_{\rm B}\right),\\ \\
\bds{\hat{\rho}} &=& \rp{\left(\bds{r}_{\rm A}-\bds{r}_{\rm B}\right)}{\rho},
\end{array}
\right.
\eqf
to be evaluated onto the unperturbed Keplerian ellipses
\eqi r_j = \rp{a_j(1-e^2_j)}{1+e_j\cos f_j},\ j={\rm A,B}\lb{ellips}\eqf
of the two test particles A and B,
it follows, for a generic perturbation \citep{cheng},
\eqi\Delta\rho=\left(\Delta\bds{r}_{\rm A}-\Delta\bds{r}_{\rm B}\right)\cdot\bds{\hat{\rho}}.\lb{range}\eqf
In it
\eqi\Delta \bds{r}_j = \Delta R_j\ \bds{\hat{R}}_j+\Delta T_j\ \bds{\hat{T}}_j+\Delta N_j\ \bds{\hat{N}}_j,\ j={\rm A,B},\lb{perti}\eqf
where $\bds{\hat{R}},\bds{\hat{T}},\bds{\hat{N}}$ are the unit vectors along the radial, transverse and out-of-plane directions, which are, in a body-centered $\{x,y,z\}$ frame, \citep{cheng}
\eqi \bds{\hat{R}} =\left(
       \begin{array}{c}
          \cos\Omega\cos u\ -\cos I\sin\Omega\sin u\\
          \sin\Omega\cos u + \cos I\cos\Omega\sin u\\
         \sin I\sin u \\
       \end{array}\lb{krizza}
     \right)
\eqf
 \eqi \bds{\hat{T}} =\left(
       \begin{array}{c}
         -\sin u\cos\Omega-\cos I\sin\Omega\cos u \\
         -\sin\Omega\sin u+\cos I\cos\Omega\cos u \\
         \sin I\cos u \\
       \end{array}
     \right)
\eqf
 \eqi \bds{\hat{N}} =\left(
       \begin{array}{c}
          \sin I\sin\Omega \\
         -\sin I\cos\Omega \\
         \cos I.\\
       \end{array}
     \right)\lb{krozza}
\eqf
In the case of the unperturbed Keplerian ellipse it is
\eqi \bds r= r\ \bds{\hat{R}},\eqf with $r$ as in \rfr{ellips} and $\bds{\hat{R}}$ given by \rfr{krizza}.
Concerning the gravitomagnetic field of the Earth, \rfr{inculescion}, evaluated for the spacecrafts A and B, has to be inserted into \rfr{range}-\rfr{perti} to yield
$\Delta\rho_{\rm LT}$; \rfr{anomazza} allows to plot it as a function of time $t$.
}
\textcolor{black}{To avoid possible misunderstandings, it is important to point out that, in order to have the correct expression for the perturbation $\Delta\rho_{\rm LT}$ of the intersatellite range, all the three components $\Delta R_{\rm LT},\Delta T_{\rm LT},\Delta N_{\rm LT}$ of \rfr{inculescion} for both the satellites A and B concur to form $\Delta\rho$ through \rfr{range} and \rfr{perti}. In other words, it would be incorrect to only consider $\Delta R_{\rm LT}$ and taking something like, say, $\Delta\rho=\Delta R_{\rm A} -\Delta R_{\rm B}$, although, at a first sight, one may be tempted to do so. }

\textcolor{black}{
Although more cumbersome, it is possible to analytically work out  the two-body range-rate perturbation $\Delta\dot\rho$ \citep{cheng} as well. The  first step consists of working out the $R-T-N$ shifts of the test-particle's velocity. In general, they are
%
\citep{Caso}
\eqi
\left\{
\begin{array}{lll}
\Delta v_R &=&    -\rp{n\sin f}{\sqrt{1-e^2}} \left( \rp{e\Delta a}{2} + \rp{a^2\Delta e}{r}\right)-\rp{n a^2\sqrt{1-e^2}}{r}\left(\cos I\Delta\Omega+\Delta\omega\right)-\rp{na^3}{r^2}\Delta \mathcal{M}, \\ \\
\Delta v_T &=& -\rp{na\sqrt{1-e^2}}{2r}\Delta a +\rp{an\left(e+\cos f\right)}{(1-e^2)^{3/2}}\Delta e +\rp{nae\sin f}{\sqrt{1-e^2}}\left(\cos I\Delta\Omega+\Delta\omega\right), \\ \\
\Delta v_N &=& \rp{na}{\sqrt{1-e^2}}\left[\left(\cos u+e\cos\omega\right)\Delta I +\left(\sin u +e \sin\omega\right)\sin I \Delta\Omega\right].
\end{array}\lb{velarr}
\right.
\eqf
In the case of the gravitomagnetic field,  the resulting velocity shifts
are
\eqi
\left\{
\begin{array}{lll}
  \Delta v_R^{\rm LT} &=& \rp{2GS\cos I\left(1+e\cos f\right)\left[2\left(f - f_0\right) -\sin\left(f - f_0\right) + e\left(\sin f-\sin f_0\right)\right]}{c^2  a^2 \left(1-e^2\right)^2},\\  \\
  \Delta v_T^{\rm LT} &=&  \rp{GS\cos I \left\{ 2\left(e + \cos f\right)\cos f_0 - \left(2 + e^2\right) + 2\left(1 + e^2\right)\sin f\sin f_0 - e \left[ 2\cos f + 4\left(f - f_0\right)\sin f -e\cos 2f\right]\right\}  }{c^2  a^2 \left(1-e^2\right)^2},\\  \\
  \Delta v_N^{\rm LT} &=&
  %
  %
  %
  %
  %
  %
  %
  %
  %
  %
  %
  \rp{GS\sin I}{c^2 a^2 \left(1-e^2\right)^2}
  \left\{
  2\left(e\cos\omega + \cos u\right)\left[\left(1 + e\cos f_0\right)\cos^2 u_0 - \left(1 + e\cos f\right)\cos^2 u\right] +\right. \\ \\
  & + &\left. \left(e\sin\omega + \sin u\right)\left[2\left(f-f_0\right) + 2e\left(\sin f-\sin f_0\right) -\left(1+e\cos f\right)\sin 2u + \right.\right. \\ \\
  & + & \left.\left.\sin 2u_0 +e\cos f_0\sin 2u_0\right]
  \right\}.
\end{array}\lb{vel_LT}
\right.
\eqf
Also \rfr{vel_LT} is an exact result in the sense that no approximations in $e$ were used.  }

\textcolor{black}{
Then, the following unit vector, computed onto the unperturbed Keplerian ellipse, is needed \citep{cheng}
\eqi \bds{\hat{\rho}}_{\nu}\doteq \rp{\left(\bds{v}_{\rm A}-\bds{v}_{\rm B}\right)-\dot\rho\ \bds{\hat{\rho}}}{\rho},\lb{vel1}\eqf
where
\eqi \dot\rho = \left(\bds{v}_{\rm A}-\bds{v}_{\rm B}\right)\cdot\bds{\hat{\rho}}.\lb{vel2}\eqf
It is intended that
the Keplerian test particle's velocity has to be used in \rfr{vel1}-\rfr{vel2}; it is
\eqi \bds{v}=v_R\ \bds{\hat{R}} + v_T\ \bds{\hat{T}},\eqf
with
\eqi
\left\{
\begin{array}{lll}
v_R &=& \rp{nae\sin f}{\sqrt{1-e^2}}, \\ \\
v_T &=& \rp{na\left(1+e\cos f\right)}{\sqrt{1-e^2}}.
\end{array}
\right.
\eqf
Note that, by construction, $\bds{\hat{\rho}}_{\nu}$ is orthogonal to $\bds{\hat{\rho}}$.
The two-body range-rate perturbation $\Delta\dot\rho$ is, thus, \citep{cheng}
\eqi \Delta\dot\rho = \left(\Delta \bds{v}_{\rm A}-\Delta \bds{v}_{\rm B}\right)\cdot \bds{\hat{\rho}} + \left(\Delta \bds{r}_{\rm A}-\Delta \bds{r}_{\rm B}\right)\cdot \bds{\hat{\rho}}_{\nu},\lb{rangerate}\eqf
where
\eqi \Delta \bds{v}_j = \Delta v_R^j\ \bds{\hat{R}}_j+\Delta v_T^j\ \bds{\hat{T}}_j+\Delta v_N^j\ \bds{\hat{N}}_j,\ j={\rm A,B}.\lb{referee1}\eqf
In the present case, inserting \rfr{inculescion} and \rfr{vel_LT} with \rfr{anomazza}, allows to plot the gravitomagnetic range-rate perturbation as a function of time. \textcolor{black}{Also in this case, we warn the reader that it would be incorrect to make, say, $\Delta\dot\rho=\Delta v_R^{\rm A} - \Delta v_R^{\rm B}$: all the terms of \rfr{vel_LT} for both satellites A and B must be used in \rfr{rangerate} through \rfr{referee1}.}
An alternative approach to compute $\Delta\dot\rho$ consists of straightforwardly taking the derivative \textcolor{black}{of} $\Delta\rho$ with respect to $t$ after that its time series has been produced\textcolor{black}{: indeed, it can be shown (see Figure \ref{figuraLT} in Section \ref{discu}) that the result is the same.}
}
\subsection{Discussion of the obtained results}\lb{discu}
In Figure \ref{figuraLT} we display the result of our numerical and analytical calculations for the gravitomagnetic field of the Earth; \textcolor{black}{Table \ref{resume1}  resumes some quantitative features of the Lense-Thirring range and range-rate signatures.}
\begin{figure*}[ht!]
\centering
\begin{tabular}{cc}
\epsfig{file=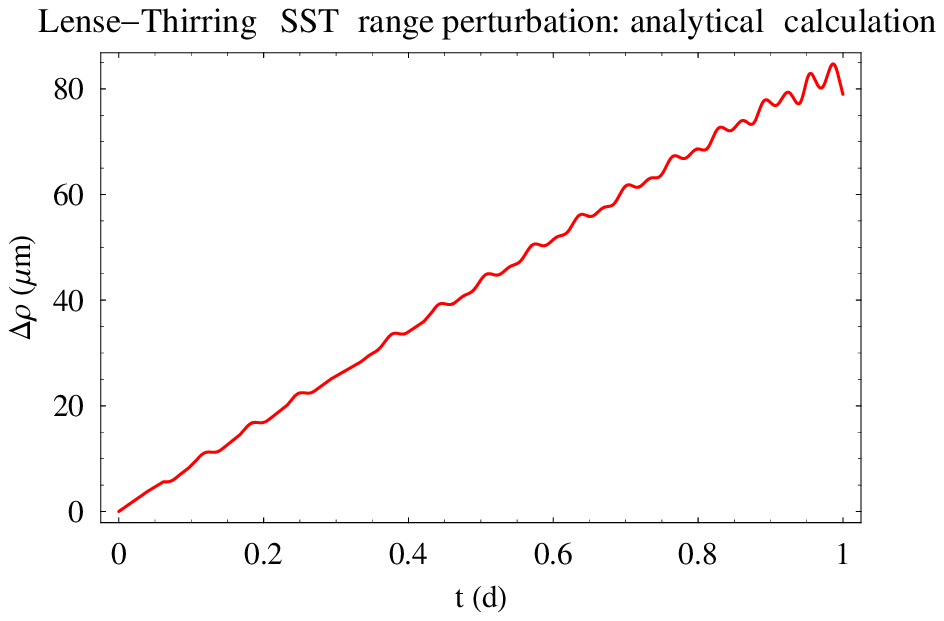,width=0.45\linewidth,clip=} &
\epsfig{file=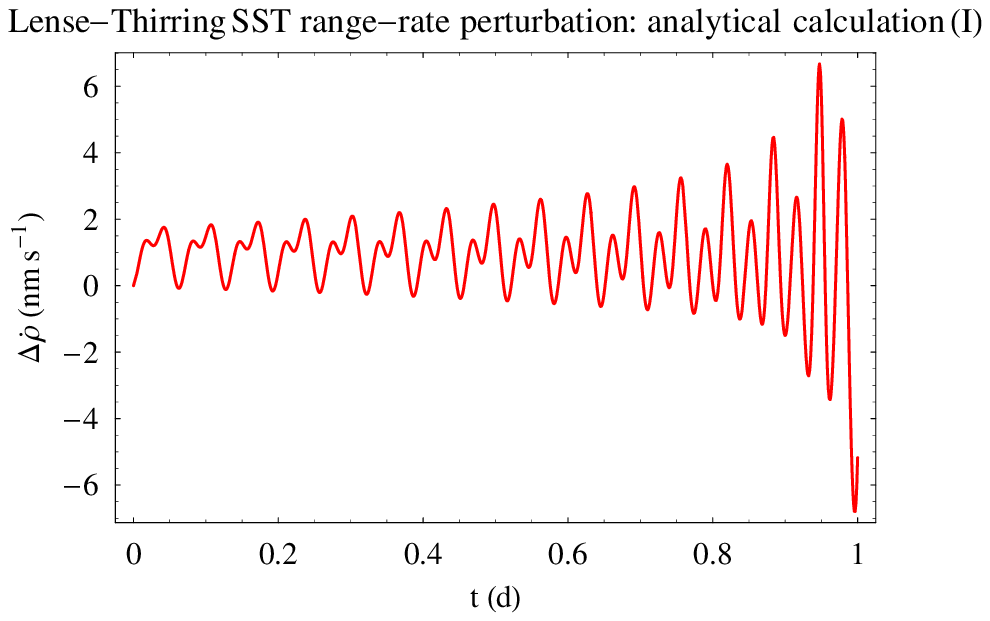,width=0.45\linewidth,clip=} \\
& \epsfig{file=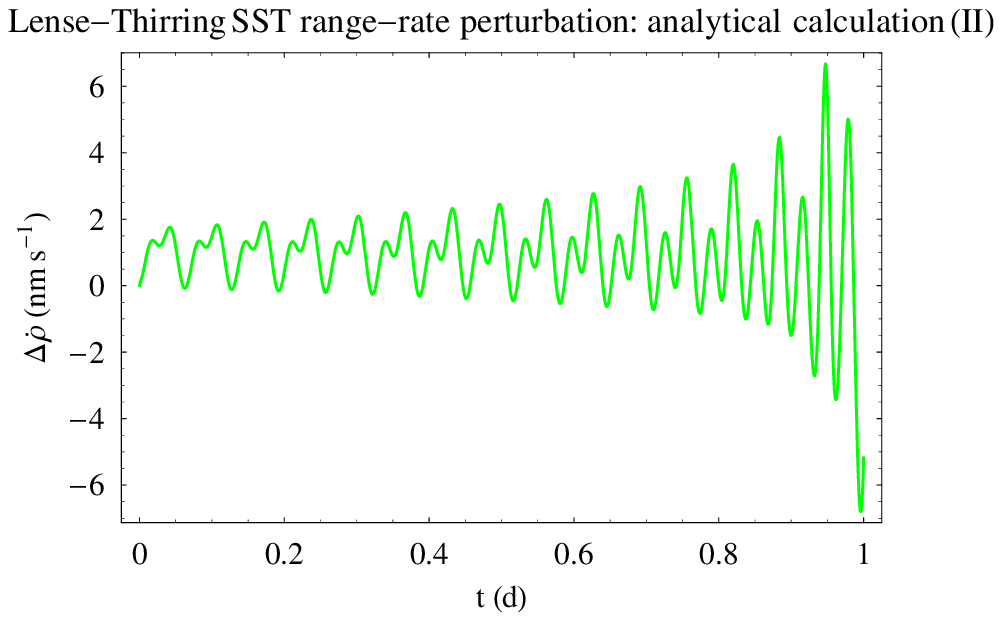,width=0.45\linewidth,clip=}\\
\epsfig{file=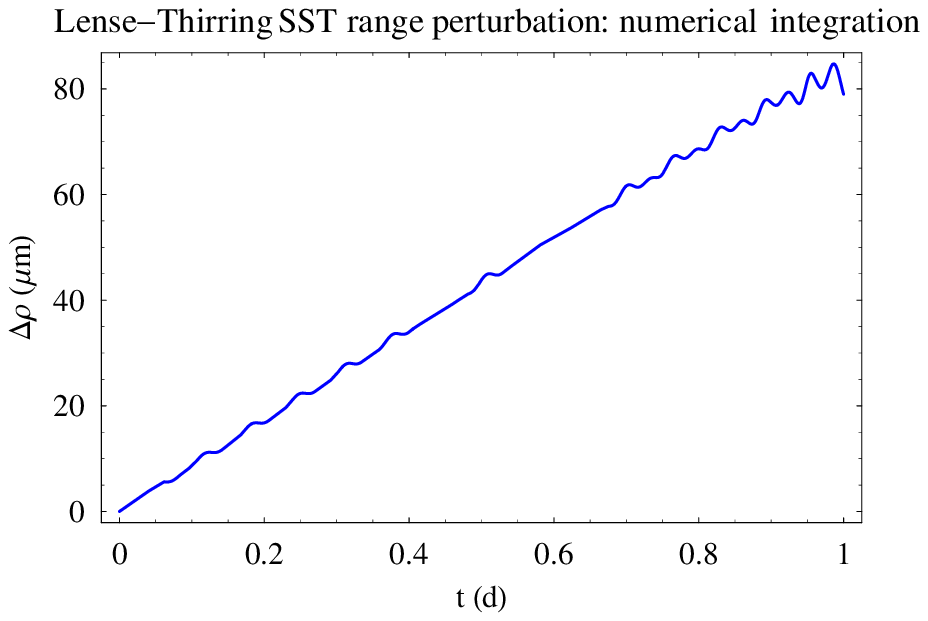,width=0.45\linewidth,clip=} &
\epsfig{file=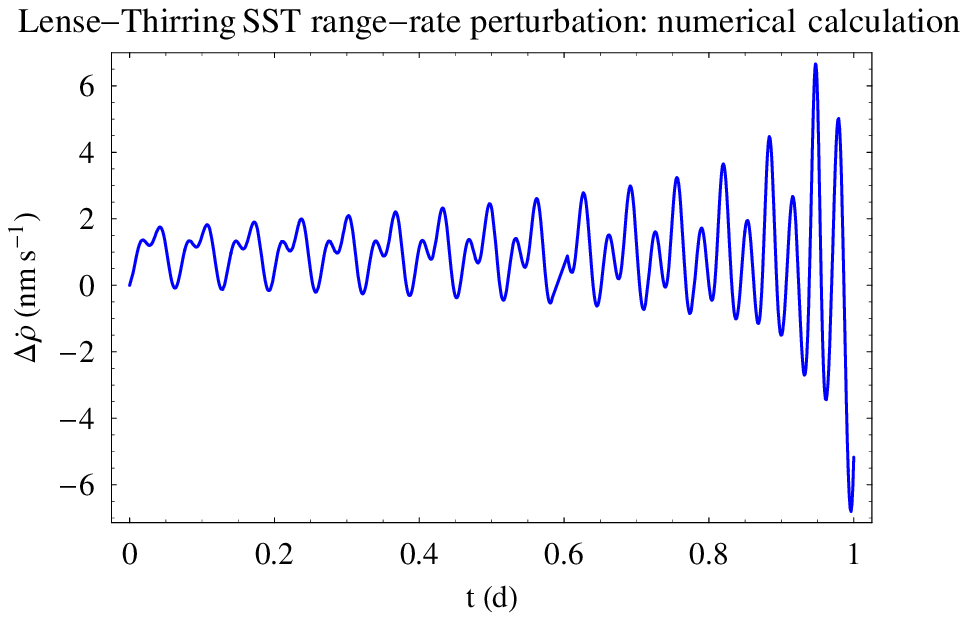,width=0.45\linewidth,clip=}
\end{tabular}
\caption{\textcolor{black}{First row from the top (red lines): analytically computed satellite-to-satellite range (left column) and range-rate (right column) signatures  for GRACE A/B due to the Lense-Thirring effect. \textcolor{black}{The range-rate signal has been obtained by differencing the range time series}.
\textcolor{black}{Second row from the top (green line): analytically computed satellite-to-satellite range-rate (right column) signature  for GRACE A/B due to the Lense-Thirring effect according to \rfr{rangerate} and \rfr{referee1} with \rfr{vel_LT} \citep{cheng}.}
Third row from the top (blue lines): differences of the numerically integrated  satellite-to-satellite range (left column) and range-rate (right column) signatures  for GRACE A/B with and without the gravitomagnetic dynamical perturbation. The initial conditions, quoted in Table \protect\ref{stateveckep},  are common to both the perturbed and unperturbed integrations. The time span is $\Delta P=1$ d. The units are $\mu$m (range) and nm s$^{-1}$  (range-rate).}}\lb{figuraLT}
\end{figure*}
\textcolor{black}{Concerning the analytical calculations for the range-rate displayed in the right-hand column of Figure \ref{figuraLT}, the red signal (I) has been obtained by differencing the corresponding red range signal on the left. Instead, the green signal (II) has been obtained from the full application of the formalism developed in Section \ref{anale} based on \rfr{vel_LT},  \rfr{rangerate} and \rfr{referee1} \citep{cheng}. As previously anticipated in Section \ref{anale}, both the approaches are equivalent. Moreover, they agree with the numerical range-rate signal as well.
At a first sight, one might find the pattern of the range-rate time series on the right somewhat puzzling if compared to that of the range ones on the left: there are intervals in the left-hand column where the plotted function is almost a straight line, while in the right-hand column its  derivative is varying as rapidly as in other intervals. Actually, there is no contradiction at all, if one carefully looks at the scales displayed on the axes of the plots. Indeed, the magnitude of the range-rate signal is of the order of a few $ {\rm n m\ s^{-1}}$, which is perfectly in agreement with the quite small variations superimposed to the linear trend of the range, plotted with $\Delta P=1\ {\rm d}=86400\ {\rm s}$. Moreover, from a  visual inspection of Figure \ref{figuraLT} it can be noticed that  the range shift amounts to 80 ${\rm \mu m}$ over 86400 s, yielding, on average, a range-rate of $\sim 0.9\ {\rm n m\ s^{-1}}$. Now, looking at the pictures in the right-hand column of Figure \ref{figuraLT}, it appears that they are not exactly symmetric with respect to the horizontal axis, with an average slightly above it. This is fully confirmed by a quantitative numerical analysis of the range-rate time series of Figure \ref{figuraLT} which yields just an average of $0.9\ {\rm n m\ s^{-1}}$. To further dispel possible doubts on the reliability of our calculations, in Figure \ref{LTtest} we plot the analytical range and range-rate time series over a shorter time span, amounting to $\Delta P = \left(P_{\rm A}^{\rm (Kep)} + P_{\rm B}^{\rm (Kep)}\right)/2=0.065\ {\rm d}=5630\ {\rm s}$, so that it is easier to see how the changes in $\Delta\rho$ are reflected by the time series of $\Delta\dot\rho$.
\begin{figure*}[ht!]
\centering
\begin{tabular}{c}
\epsfig{file=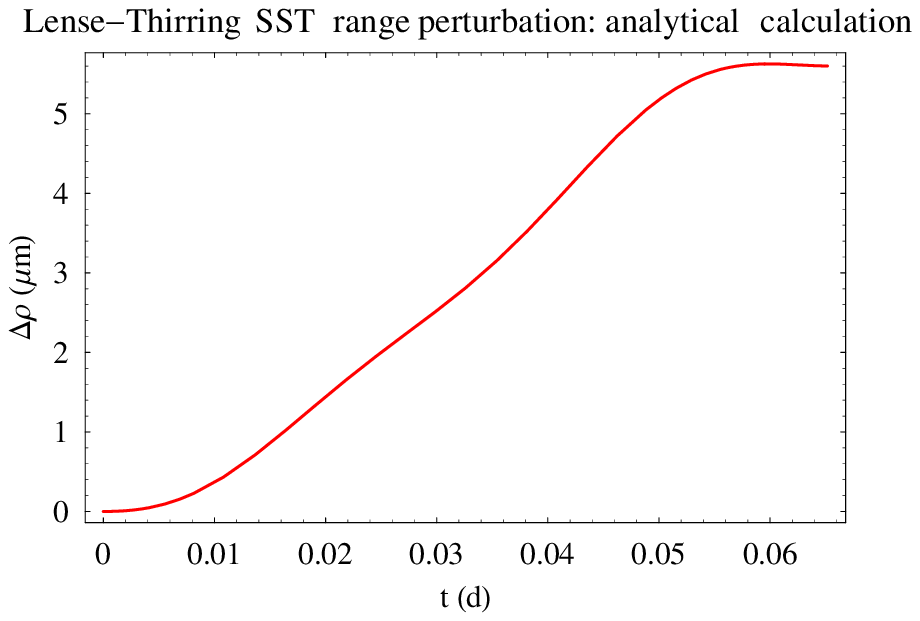,width=0.45\linewidth,clip=} \\
\epsfig{file=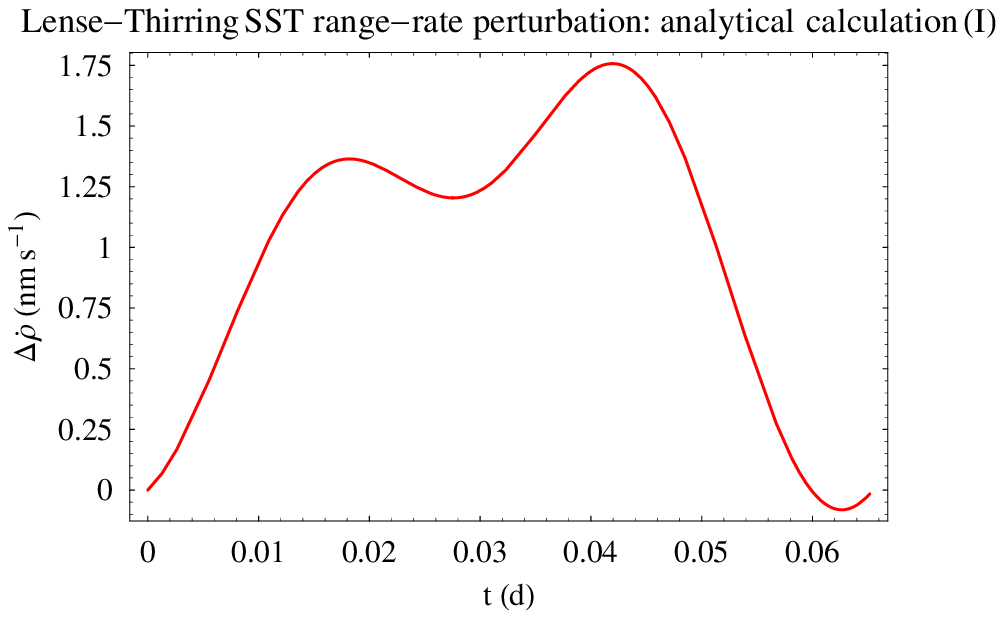,width=0.45\linewidth,clip=}
\end{tabular}
\caption{\textcolor{black}{First panel from the top: analytically computed satellite-to-satellite range  signature, in $\mu$m,  for GRACE A/B due to the Lense-Thirring effect.
Second panel from the top: analytically computed satellite-to-satellite range-rate signature, in nm s$^{-1}$,  for GRACE A/B due to the Lense-Thirring effect.
The details of the calculation are as in Figure \ref{figuraLT}. The time span of the plots is $\Delta P=0.065\ {\rm d}$.}
}\lb{LTtest}
\end{figure*}
}

As far as Figure \ref{figuraLT} is concerned, the magnitude of the Lense-Thirring range signal is $\Delta\rho_{\rm LT}=80$ $\mu$m, while the gravitomagnetic range-rate effect is quite small, being of the order of $\Delta\dot\rho_{\rm LT}=0.012$ $\mu$m s$^{-1}$ (peak-to-peak amplitude).
By assuming a present-day accuracy of $\sigma_{\rho}\sim 5\ \mu$m in measuring the satellite-to-satellite range, the Lense-Thirring signature would fall, in principle, within the measurability domain at a $\sim 6\%$ level. Concerning the range-rate, if we assume $\sigma_{\dot\rho}\sim 0.5\ \mu$m s$^{-1}$,  the Lense-Thirring effect on the range-rate $\Delta\dot\rho_{\rm LT}$ is, instead, certainly too small to be detectable.

Moving to the largest general relativistic orbital perturbation, i.e. the one due the gravitoelectric, Schwarzschild-like part of the gravitational field, Figure \ref{figuraGE} depicts its numerically integrated range and range-rate signatures. \textcolor{black}{Table \ref{resume1} summarizes some quantitative features of the gravitoelectric perturbations.}
\begin{figure*}[ht!]
\centering
\begin{tabular}{cc}
\epsfig{file=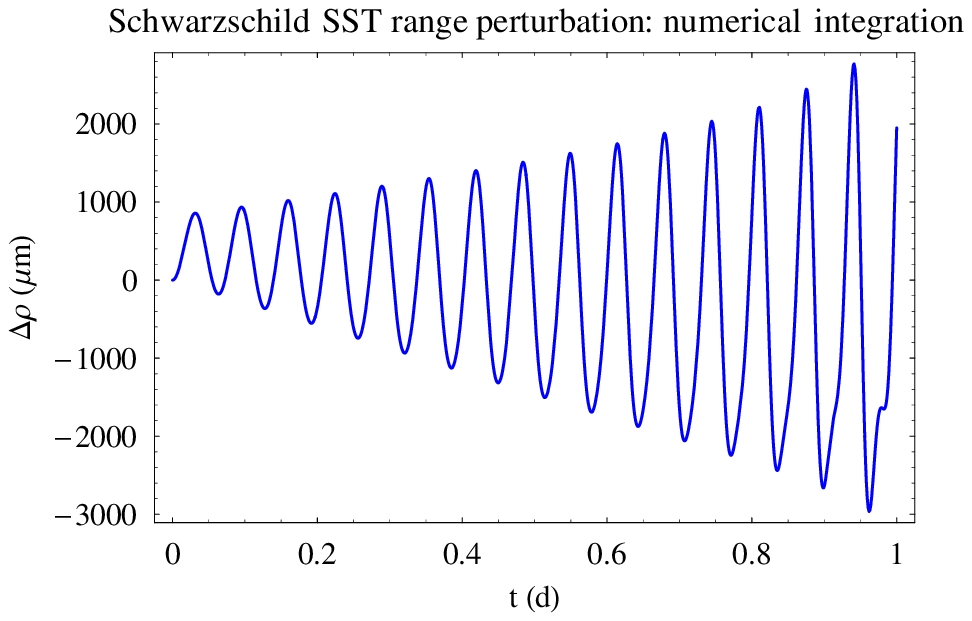,width=0.45\linewidth,clip=} &
\epsfig{file=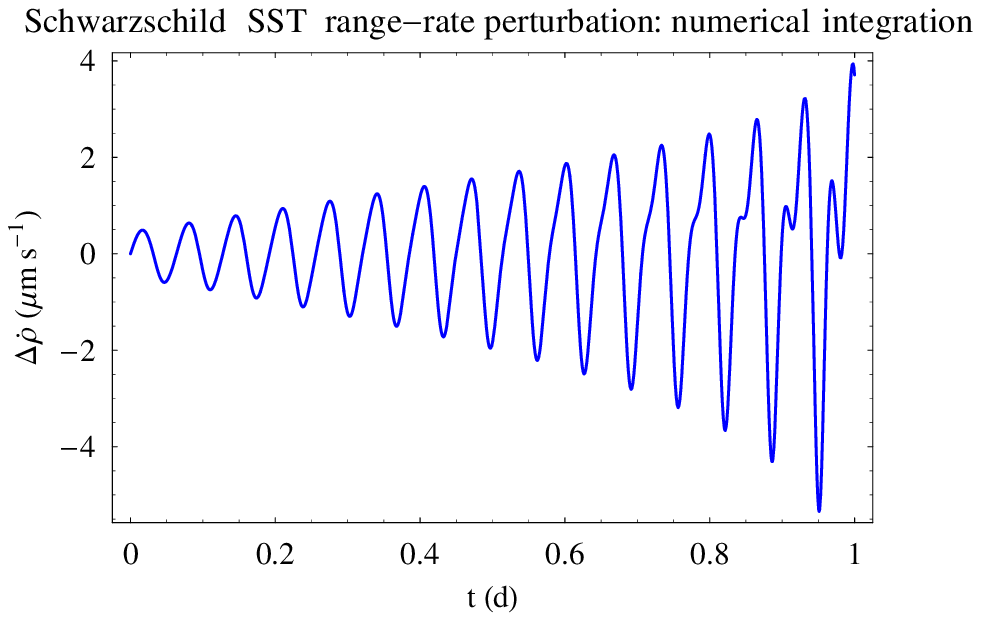,width=0.45\linewidth,clip=}
\end{tabular}
\caption{\textcolor{black}{Differences of the numerically integrated  satellite-to-satellite range (left column) and range-rate (right column) signatures  for GRACE A/B with and without the Schwarzschild dynamical perturbation. The initial conditions, quoted in Table \protect\ref{stateveckep},  are common to both the perturbed and unperturbed integrations. The time span is $\Delta P=1$ d. The units are $\mu$m (range) and $\mu$m s$^{-1}$  (range-rate).}}\lb{figuraGE}
\end{figure*}
The peak-to-peak amplitude of the range effect is as large as $\Delta\rho_{\rm Schw}=6000$ $\mu$m, so that it would be, in principle, detectable at a $8\times 10^{-4}$ level. The range-rate signal amounts to about $\Delta\dot\rho_{\rm Schw}=10$ $\mu$m s$^{-1}$, which would be measurable with a nominal accuracy of about $5\%$.

Thus, GTR affects the orbital part of the GRACE satellite-to-satellite range in a detectable way, at least in principle, given the present-day level of accuracy in measuring it. The same also holds for the range-rate, although only for the Schwarzschild signal, and at a lower level of accuracy with respect to the range.
\begin{table*}[ht!]
\caption{Peak-to-peak amplitudes of the inter-satellite GRACE range and range-rate perturbations $\Delta\rho$, $\Delta\dot\rho$ caused by the general relativistic Schwarzschild and Lense-Thirring components \textcolor{black}{of the Earth's gravitational field.}
The present-day accuracy in measuring the satellite-to satellite GRACE range and range-rate is   $\sigma_{\rho}\lesssim 10\ \mu$m and  $\sigma_{\dot\rho}\lesssim 1\ \mu$m s$^{-1}$, respectively.
}\label{resume1}
\centering
\bigskip
\begin{tabular}{lll}
\hline\noalign{\smallskip}
Dynamical effect & $\Delta\rho$ ($\mu$m)& $\Delta\dot\rho$ ($\mu$m s$^{-1}$)  \\
\noalign{\smallskip}\hline\noalign{\smallskip}
Schwarzschild & $6000$ & $10$\\
Lense-Thirring & $80$ & $0.012$ \\
%
%
%
%
%
%
%
\noalign{\smallskip}\hline\noalign{\smallskip}
\end{tabular}
\end{table*}
However, we wish to point out that it should, actually, be  checked if the relativistic signatures are not  absorbed and removed from the range signal in estimating some of the various range parameters which are solved-for in the usual GRACE data processing. Indeed, it is exposed to some mismodeled device behavior, which requires estimating many empirical parameters in semi-dynamical orbit processing mode \citep{bias}. In fact, it would be necessary to realistically simulate the range and range-rate observations by fully modeling GTR, and implementing a data reduction: it is beyond the scope of the present work. On the other hand, our choice for the time interval of the numerical integrations, close to the ones actually employed in real data reductions \citep{Flech011}, makes our analysis closer to what is actually done in real GRACE data analyses.

In Section \ref{terza} we will look at the competing classical effects induced on the range and range-rate of GRACE by some low-degree zonal harmonics $J_{\ell}, \ell\geq 2$ of the multipolar expansion of the Newtonian part of the Earth's gravitational potential accounting for its departures from spherical symmetry. Indeed, their unavoidably imperfect knowledge causes  mismodeled range and range-rate signals which would corrupt the recovery of the relativistic ones at a level which has to be quantitatively assessed. On the other hand, such an investigation will also contribute to yield quantitative, although preliminary, evaluations of  the level of a possible a-priori \virg{imprinting} of GTR itself, not solved-for so far in all the GRACE-based models, on the estimated values of such zonals. This issue was treated, in the framework of the LAGEOS-based tests of the Lense-Thirring effect, by \citet{CN} as far as the node precessions of the orbital planes of the GRACE spacecrafts are concerned.
\section{A-priori \virg{imprint} level of GTR in the  zonal KBR signature}\lb{terza}
In this Section we numerically work out the effects on the GRACE range and  range-rate  caused by the mismodelling in the first \textcolor{black}{six} zonal coefficients of the terrestrial gravitational field. See Table \ref{zona} for \textcolor{black}{their values and} formal, statistical 1-$\sigma$ errors in one of the most recent global Earth's gravity field solution.
\begin{table*}[ht!]
\caption{Estimated values $\overline{C}_{\ell,0}$ and formal, statistical errors $\sigma_{\overline{C}_{\ell,0}}$ of the normalized Stokes coefficients of the geopotential for $\ell=2,3,4,5,6,7$ from the GOCE/GRACE-based solution GOCO01S \protect\citep{goco}. Recall that $J_{\ell}\doteq -\sqrt{2\ell+1}\ \overline{C}_{\ell,0}$.
}\label{zona}
\centering
\bigskip
\begin{tabular}{lll}
\hline\noalign{\smallskip}
Degree $\ell$ & $\overline{C}_{\ell,0}$ & $\sigma_{\overline{C}_{\ell,0}}$ \\
\noalign{\smallskip}\hline\noalign{\smallskip}
2 & $-4.841649689\times 10^{-4}$ & $4\times 10^{-13}$\\
3 & $9.571980\times 10^{-7}$ & $3\times 10^{-13}$ \\
4 & $5.4000331\times 10^{-7}$ & $8\times 10^{-14}$ \\
5 & $6.867018\times 10^{-8}$ & $6\times 10^{-14}$ \\
6 & $-1.4995817\times 10^{-7}$ & $4\times 10^{-14}$ \\
7 & $9.051062\times 10^{-8}$ & $4\times 10^{-14}$ \\
\noalign{\smallskip}\hline\noalign{\smallskip}
\end{tabular}
\end{table*}
\subsection{The even zonals}
The range and range-rate perturbations due to the  mismodelling in the first three even zonals are depicted in Figure \ref{figura2} and \textcolor{black}{quantitatively} summarized in Table \ref{Even_Zon_Tab}.
\begin{figure*}[ht!]
\centering
\begin{tabular}{cc}
\epsfig{file=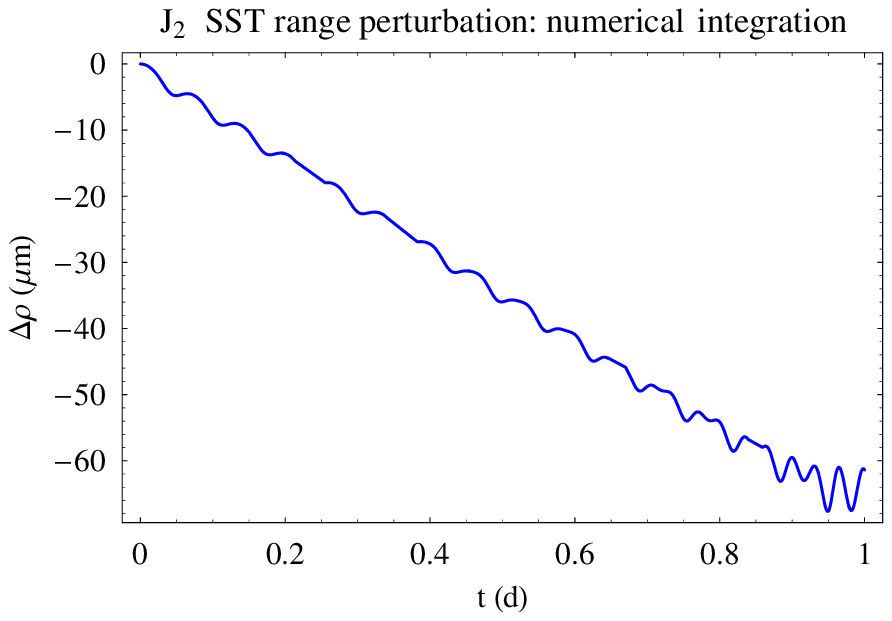,width=0.45\linewidth,clip=} &
\epsfig{file=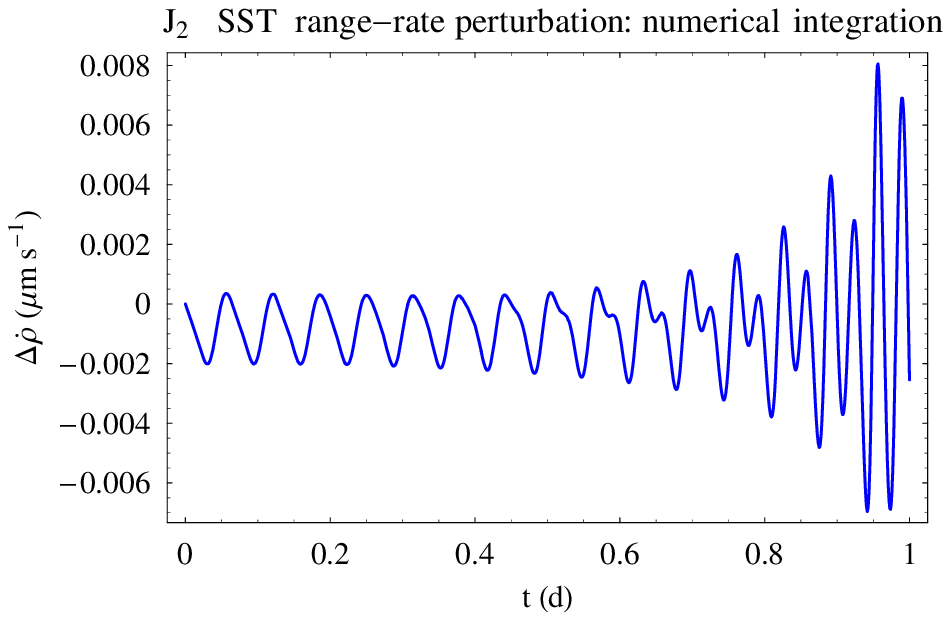,width=0.45\linewidth,clip=} \\
\epsfig{file=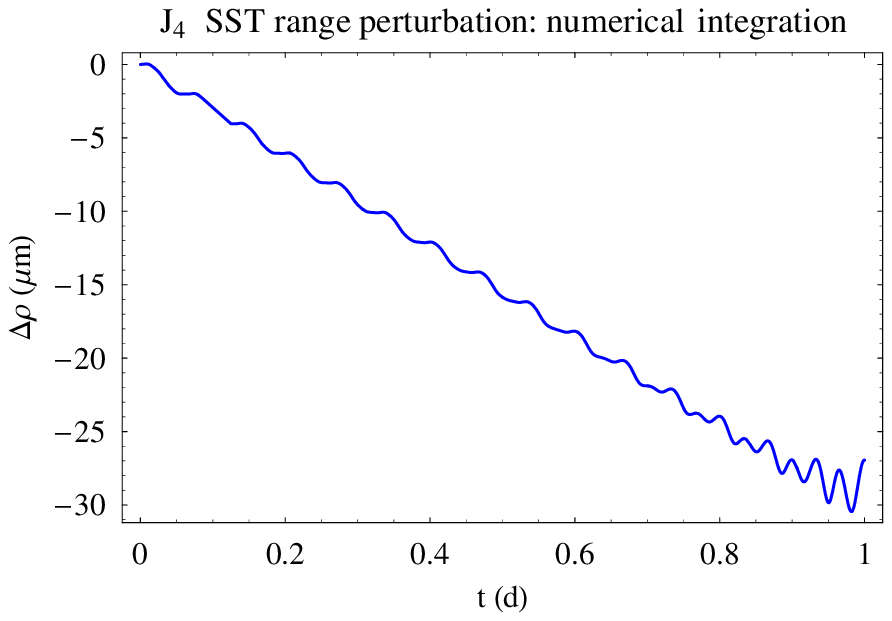,width=0.45\linewidth,clip=} &
\epsfig{file=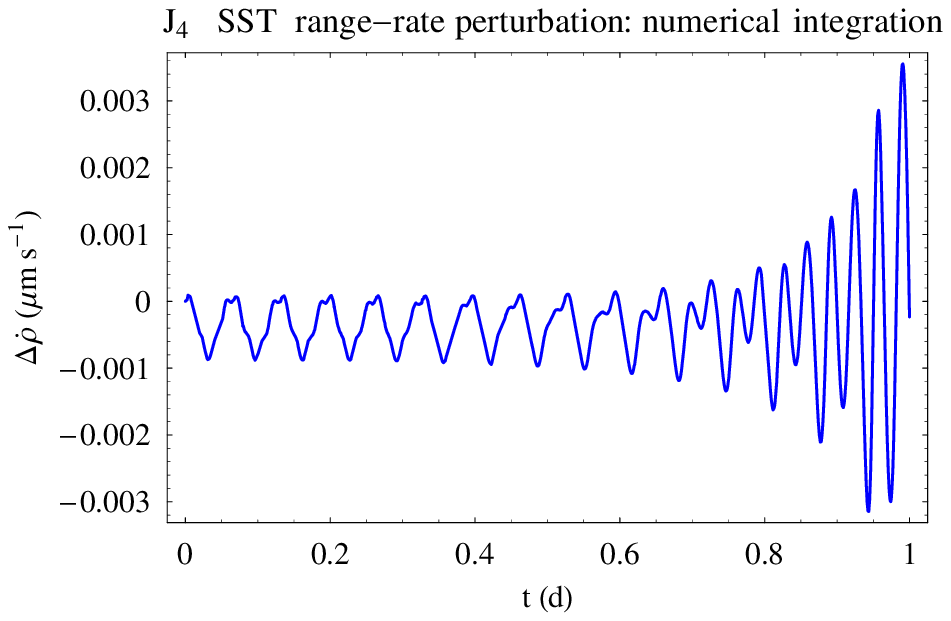,width=0.45\linewidth,clip=} \\
\epsfig{file=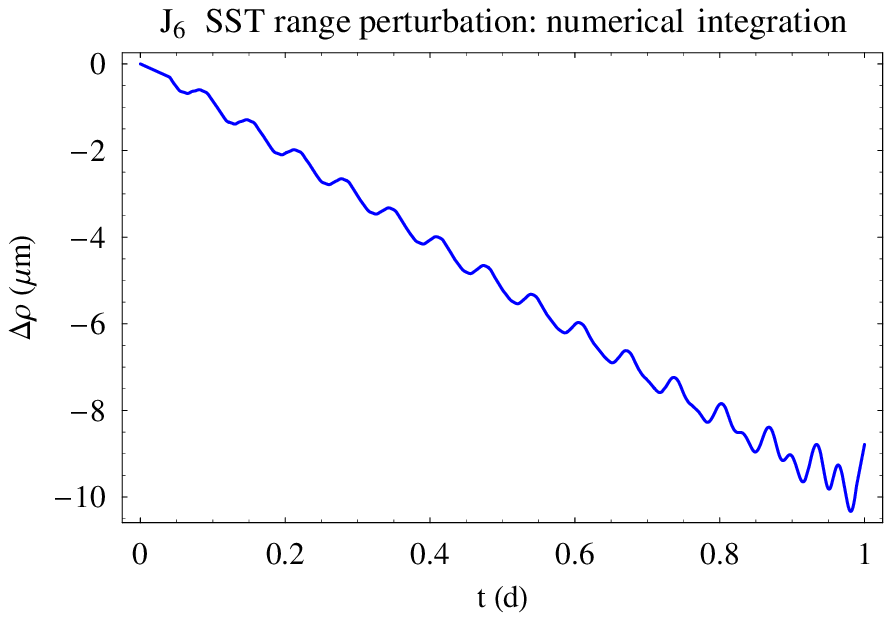,width=0.45\linewidth,clip=} &
\epsfig{file=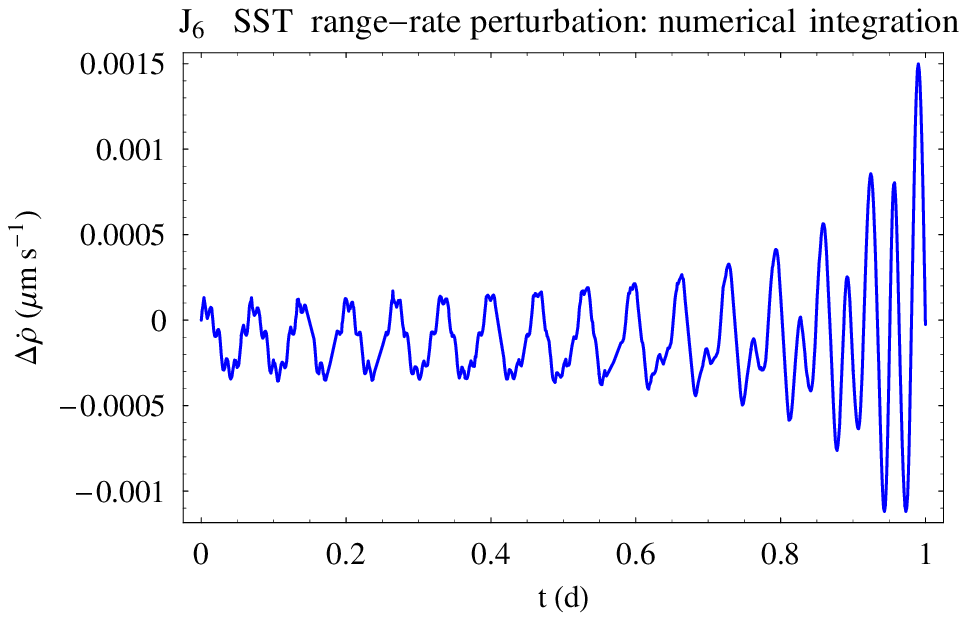,width=0.45\linewidth,clip=}
\end{tabular}
\caption{Differences of the numerically integrated  ranges (left column) and range-rates (right column) for GRACE A/B with and without the  classical  dynamical perturbations due to the mismodelled even zonal harmonics $J_{\ell}\doteq -\sqrt{2\ell +1}\ \overline{C}_{\ell, 0},\ \ell=2,4,6$.  According to the recent combined GOCE-GRACE solution GOCO01S \protect\citep{goco}, the formal, statistical uncertainties in the normalized Stokes coefficients are, $\sigma_{\overline{C}_{2,0}}=0.44\times 10^{-12}, \sigma_{\overline{C}_{4,0}}=0.8\times 10^{-13}, \sigma_{\overline{C}_{6,0}}=0.4\times 10^{-13}$ respectively. The initial conditions, quoted in Table \protect\ref{stateveckep},  are common to both the perturbed and unperturbed integrations. The time span is $\Delta P=1$ d. The units are $\mu$m (range) and $\mu$m s$^{-1}$  (range-rate).}\lb{figura2}
\end{figure*}
\begin{table*}[ht!]
\caption{\textcolor{black}{Peak-to-peak amplitudes of the inter-satellite GRACE range and range-rate perturbations $\Delta\rho$, $\Delta\dot\rho$ caused by the first three mismodelled even zonal harmonics of the classical part of the geopotential over $\Delta P=1$ d. The figures for  $\sigma_{\overline{C}_{\ell,0}},\ \ell=2,4,6$ along with their nominal values, can be found in Table \ref{zona}.}
}\label{Even_Zon_Tab}
\centering
\bigskip
\begin{tabular}{lll}
\hline\noalign{\smallskip}
Dynamical effect & $\Delta\rho$ ($\mu$m)& $\Delta\dot\rho$ ($\mu$m s$^{-1}$)  \\
\noalign{\smallskip}\hline\noalign{\smallskip}
$J_2$  & $70$ & $0.014$ \\
%
%
$J_4$  & $30$ & $0.006$ \\
%
%
$J_6$  & $10$ & $0.0025$ \\
%
%
\noalign{\smallskip}\hline\noalign{\smallskip}
\end{tabular}
\end{table*}

\textcolor{black}{According to Table \ref{resume1} and Table \ref{Even_Zon_Tab}, the mismodelled ranges due to the even zonals are slightly smaller than the corresponding Lense-Thirring effect, while they are about $90-600$ times smaller than the Schwarzschild range; instead, the Schwarzschild range-rate signal is $700-4000$ times larger than the corresponding mismodelled signatures due to the even zonals.}
However, it must be considered that, given a specific Earth's gravity field model, the actual uncertainties in its estimated even zonals may be up to one order of magnitude larger \textcolor{black}{with respect to their formal, statistical errors}. Moreover, an even more conservative approach to realistically evaluate the true uncertainties in the even zonals consists of comparing their estimated values from different global gravity field solutions \citep{Wagner011}.

Conversely, we can use \textcolor{black}{Table \ref{resume1} and Table \ref{Even_Zon_Tab} (or, equivalently, Figure \ref{figuraLT}-\ref{figuraGE} and  Figure \ref{figura2})} to obtain preliminary, quantitative evaluations of a possible a-priori \virg{imprinting} of GTR itself in the even zonals considered. \textcolor{black}{By posing $x\doteq \rho,\dot\rho$, it can be done from}
\eqi\Delta x_{\rm GTR} = \Delta \xi_{\ell} \Delta J^{\rm (eff)}_{\ell},\ \Delta\xi_{\ell}\doteq\rp{\Delta x_{J_{\ell}}}{J_{\ell}} .\eqf Thus,
by dividing the relativistic range and range-rate perturbations \textcolor{black}{$\Delta x_{\rm GTR}$} by the corresponding \textcolor{black}{normalized} classical ones \textcolor{black}{$\Delta \xi_{\ell}$} for each degree $\ell$ considered   gives us a sort of \virg{effective} relativistic even zonal $\Delta J_{\ell}^{\rm (eff)}$, i.e. the part of the even zonal of degree $\ell$ which would give a signal as large as those due to GTR.

\begin{table*}[ht!]
\caption{\textcolor{black}{\virg{Effective} general relativistic  parts  $\Delta J^{(\rm eff)}_{\ell}$  of the even zonals of degree $\ell=2,4,6$ obtained from the  range perturbations $\Delta\rho$. They are a preliminary measure of a possible a-priori \virg{imprinting} of GTR itself, not explicitly solved-for  in all the GRACE-based solutions produced so far, on the even zonals. Compare these figures with the formal, statistical errors $\sigma_{\overline{C}_{\ell,0}},\ \ell=2,4,6$ in Table \ref{zona}.}
}\label{resume2}
\centering
\bigskip
\begin{tabular}{llll}
\hline\noalign{\smallskip}
Dynamical effect ($\Delta\rho$) & $\Delta J^{(\rm eff)}_2$  & $\Delta J_4^{(\rm eff)}$ &  $\Delta J_6^{(\rm eff)}$  \\
\noalign{\smallskip}\hline\noalign{\smallskip}
Schwarzschild & $3\times 10^{-11}$ & $2\times 10^{-11}$ & $2\times 10^{-11}$ \\
Lense-Thirring & $5\times 10^{-13}$ & $2\times 10^{-13}$ & $3\times 10^{-13}$\\
\noalign{\smallskip}\hline\noalign{\smallskip}
\end{tabular}
\end{table*}
\begin{table*}[ht!]
\caption{\textcolor{black}{\virg{Effective} general relativistic  parts  $\Delta J^{(\rm eff)}_{\ell}$   of the even zonals of degree $\ell=2,4,6$ obtained from the range-rate perturbations $\Delta\dot\rho$. They are another preliminary measure of a possible a-priori \virg{imprinting} of GTR itself, not explicitly solved-for  in all the GRACE-based solutions produced so far, on the even zonals. Compare these figures with the formal, statistical errors $\sigma_{\overline{C}_{\ell,0}},\ \ell=2,4,6$ in Table \ref{zona}.}
}\label{resume3}
\centering
\bigskip
\begin{tabular}{llll}
\hline\noalign{\smallskip}
Dynamical effect ($\Delta\dot\rho$) & $\Delta J^{(\rm eff)}_2$  & $\Delta J_4^{(\rm eff)}$ &  $\Delta J_6^{(\rm eff)}$  \\
\noalign{\smallskip}\hline\noalign{\smallskip}
Schwarzschild & $3\times 10^{-10}$ & $1\times 10^{-10}$ & $2\times 10^{-10}$ \\
Lense-Thirring & $3\times 10^{-13}$ & $2\times 10^{-13}$ & $2\times 10^{-13}$\\
\noalign{\smallskip}\hline\noalign{\smallskip}
\end{tabular}
\end{table*}

According to Table \ref{resume2} and Table \ref{resume3}, and in view of the present-day level of accuracy in estimating them (see Table \ref{zona}), the a-priori \virg{imprinting} of GTR on the even zonals should not be neglected, especially as far as the Schwarzschild part is concerned. Following the approach by \citet{CN}, it can be shown that a Lense-Thirring
\virg{imprint} in $J_4$ and $J_6$ as large as that of Table \ref{resume2}-Table \ref{resume3}  would correspond to a $\approx 0.2\%$ systematic
bias in the LAGEOS-based tests of such a relativistic
effect. \textcolor{black}{However, there is room for a \virg{contamination} of GTR itself due to the gravitoelectric part.}
\subsection{The odd zonals}
%
In Figure \ref{figura3} and Table \ref{Odd_Zon_Tab} we repeat the same analysis for the first three odd zonals.
\begin{figure*}[ht!]
\centering
\begin{tabular}{cc}
\epsfig{file=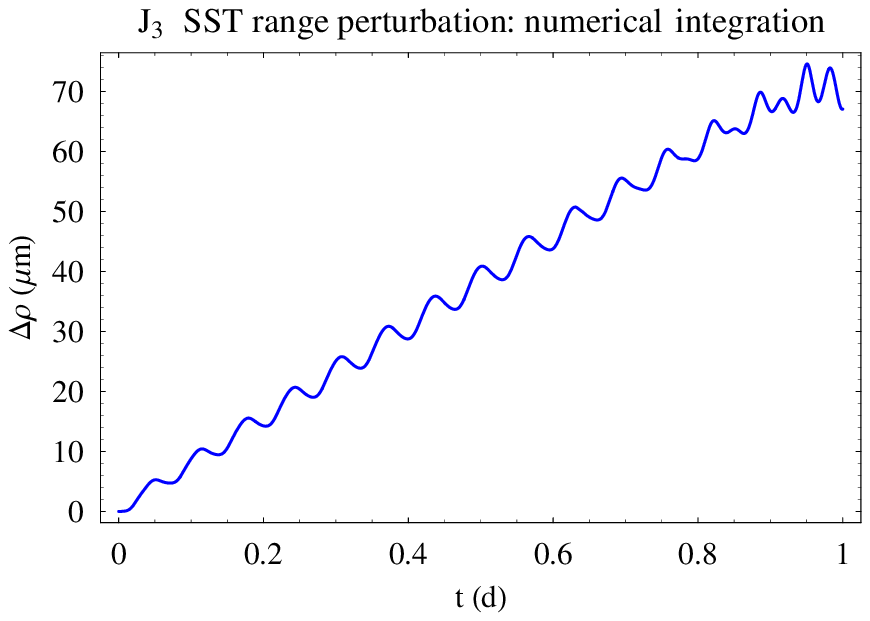,width=0.45\linewidth,clip=} &
\epsfig{file=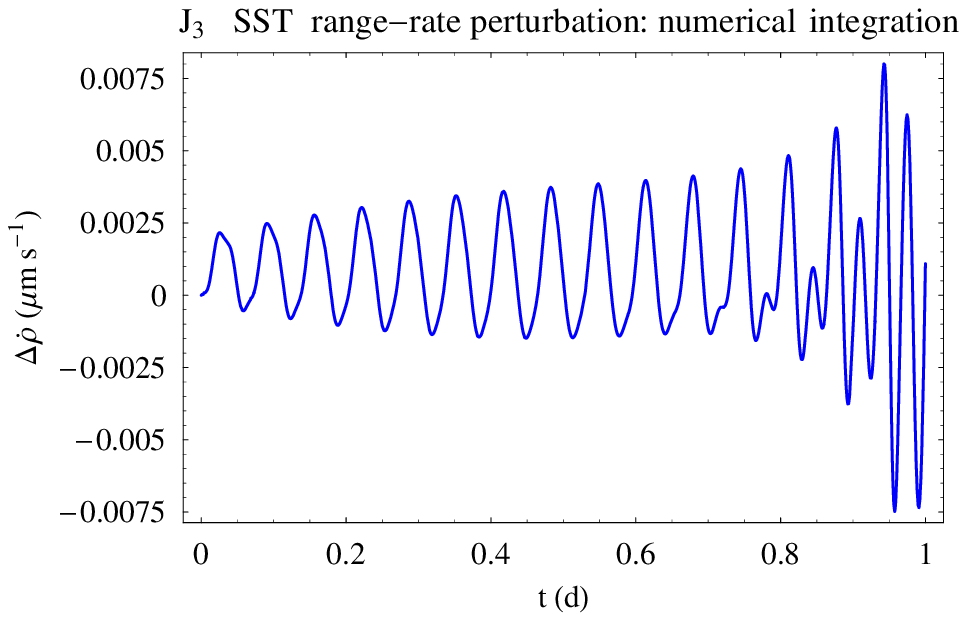,width=0.45\linewidth,clip=} \\
\epsfig{file=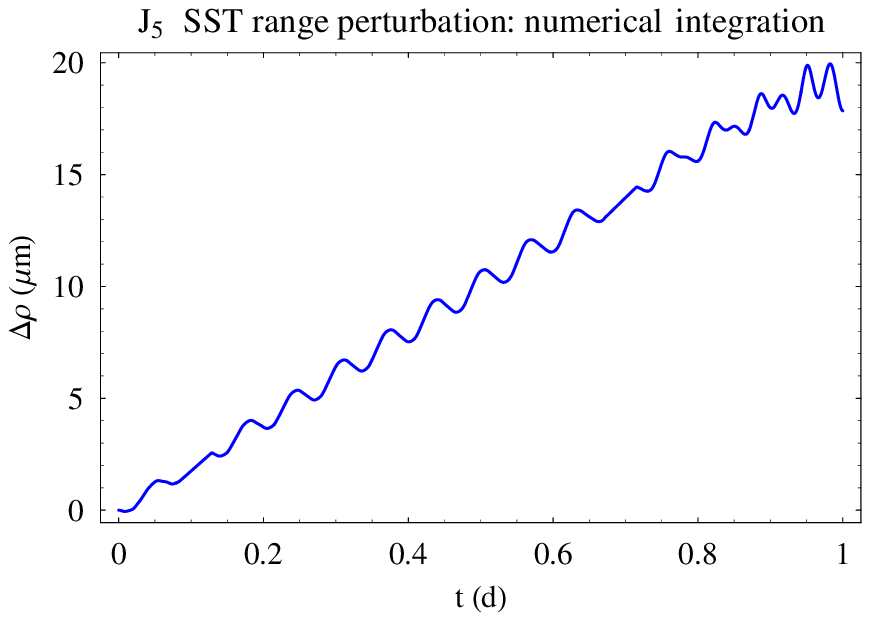,width=0.45\linewidth,clip=} &
\epsfig{file=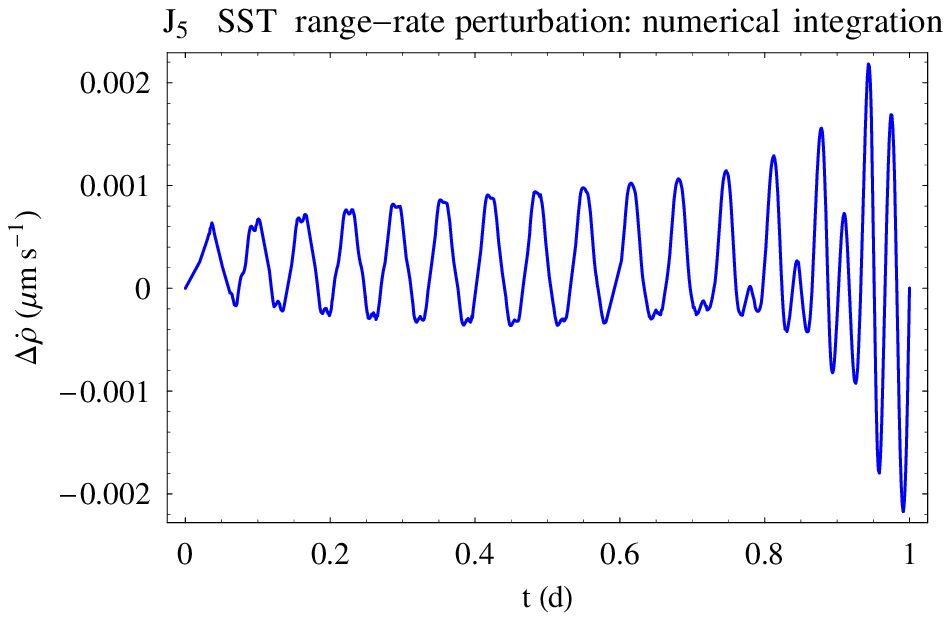,width=0.45\linewidth,clip=} \\
\epsfig{file=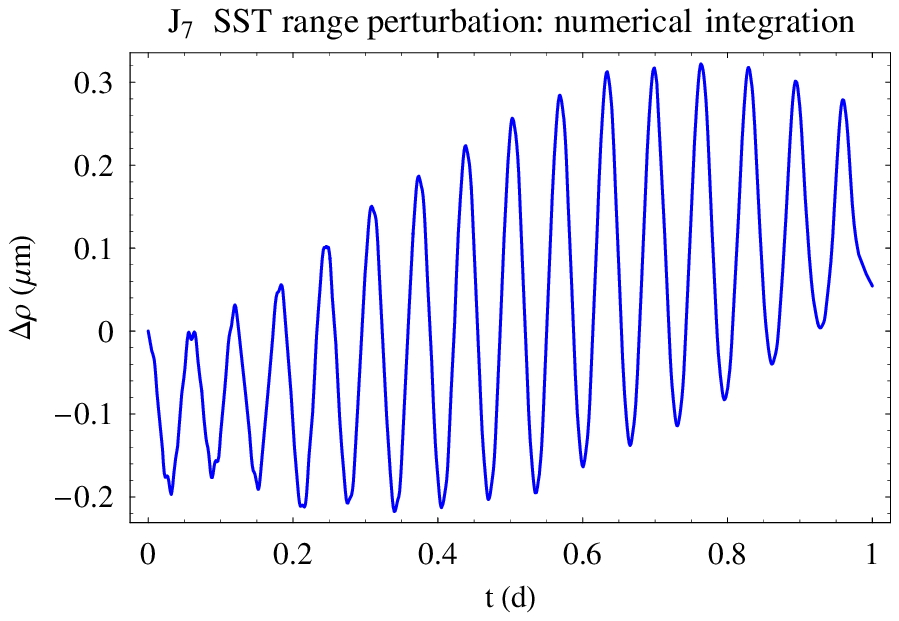,width=0.45\linewidth,clip=} &
\epsfig{file=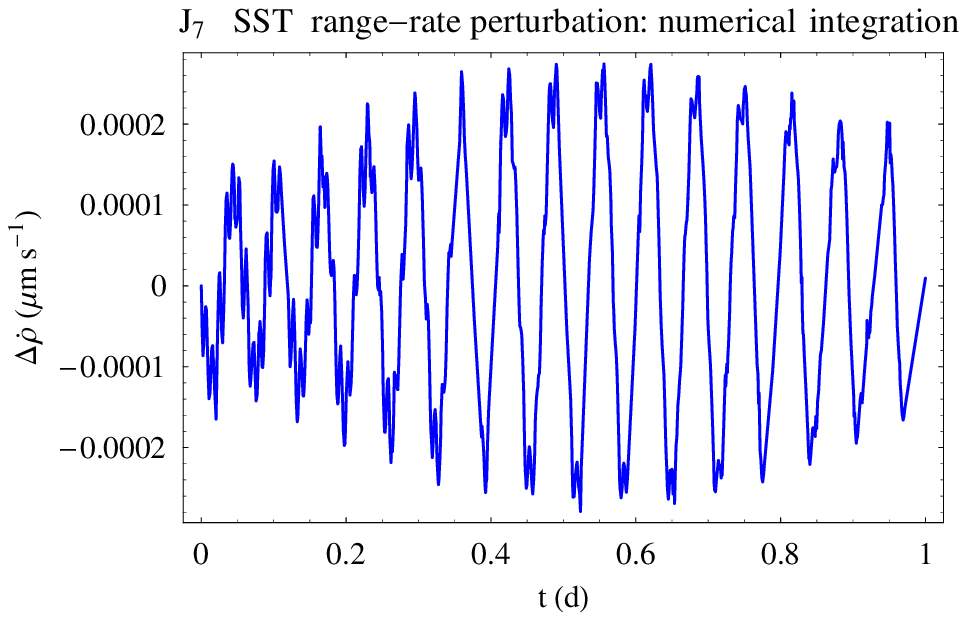,width=0.45\linewidth,clip=}
\end{tabular}
\caption{Differences of the numerically integrated  ranges (left column) and range-rates (right column) for GRACE A/B with and without the classical  dynamical perturbations due to the mismodelled odd zonal harmonics $J_{\ell}\doteq -\sqrt{2\ell +1}\ \overline{C}_{\ell, 0},\ \ell=3,5,7$. According to the recent combined GOCE-GRACE solution GOCO01S \protect\citep{goco}, the formal, statistical uncertainties in the normalized Stokes coefficients are, $\sigma_{\overline{C}_{3,0}}=0.27\times 10^{-12}, \sigma_{\overline{C}_{5,0}}=0.6\times 10^{-13}, \sigma_{\overline{C}_{7,0}}=0.4\times 10^{-13}$ respectively. The initial conditions, quoted in Table \protect\ref{stateveckep},  are common to both the perturbed and unperturbed integrations. The time span is $\Delta P=1$ d. The units are $\mu$m (range) and $\mu$m s$^{-1}$  (range-rate).}\lb{figura3}
\end{figure*}
\begin{table*}[ht!]
\caption{\textcolor{black}{Peak-to-peak amplitudes of the inter-satellite GRACE range and range-rate perturbations $\Delta\rho$, $\Delta\dot\rho$ caused by the first three mismodelled odd zonal harmonics of the classical part of the geopotential over $\Delta P=1$ d. The figures for  $\sigma_{\overline{C}_{\ell,0}},\ \ell=3,5,7$ along with their nominal values, can be found in Table \ref{zona}.}
}\label{Odd_Zon_Tab}
\centering
\bigskip
\begin{tabular}{lll}
\hline\noalign{\smallskip}
Dynamical effect & $\Delta\rho$ ($\mu$m)& $\Delta\dot\rho$ ($\mu$m s$^{-1}$)  \\
\noalign{\smallskip}\hline\noalign{\smallskip}
$J_3$  & $70$ & $0.015$ \\
%
%
$J_5$  & $20$ & $0.004$ \\
%
%
$J_7$  & $0.5$ & $0.0004$ \\
%
%
\noalign{\smallskip}\hline\noalign{\smallskip}
\end{tabular}
\end{table*}

\textcolor{black}{Concerning the range, it turns out that the shifts due to the mismodelled odd zonals are about as large as the Lense-Thirring signature and smaller than the Schwarzschild one by a factor $\approx 85-12000$. The  gravitomagnetic range-rate perturbation is of the same order of magnitude of the $J_3-$induced range-rate mismodelled signature, but it is larger than the other ones up to 160 times ($J_7$). The Schwarzschild range-rate effect is larger than the competing odd zonals ones by a factor $\approx 670-25000$. The Lense-Thirring range-rate signal is smaller than the mismodelled $J_3$ one, but it is $3-30$ times larger than the $J_5$ and $J_7$ mismodelled signatures.}

\textcolor{black}{The a-priori \virg{imprinting} of GTR on the odd zonals can preliminarily be quantified according to Table \ref{imprin1} (range) and Table \ref{imprin2} (range-rate).
\begin{table*}[ht!]
\caption{\textcolor{black}{\virg{Effective} general relativistic  parts  $\Delta J^{(\rm eff)}_{\ell}$  of the odd zonals of degree $\ell=3,5,7$ obtained from the range perturbations $\Delta\rho$. They are a preliminary measure of a possible a-priori \virg{imprinting} of GTR itself, not explicitly solved-for  in all the GRACE-based solutions produced so far, on the odd zonals. Compare these figures with the formal, statistical errors $\sigma_{\overline{C}_{\ell,0}},\ \ell=3,5,7$ in Table \ref{zona}.}
}\label{imprin1}
\centering
\bigskip
\begin{tabular}{llll}
\hline\noalign{\smallskip}
Dynamical effect ($\Delta\rho$) & $\Delta J^{(\rm eff)}_3$  & $\Delta J_5^{(\rm eff)}$ &  $\Delta J_7^{(\rm eff)}$  \\
\noalign{\smallskip}\hline\noalign{\smallskip}
Schwarzschild & $3\times 10^{-11}$ & $2\times 10^{-11}$ & $5\times 10^{-10}$ \\
Lense-Thirring & $4\times 10^{-13}$ & $2\times 10^{-13}$ & $6.4\times 10^{-12}$\\
\noalign{\smallskip}\hline\noalign{\smallskip}
\end{tabular}
\end{table*}
\begin{table*}[ht!]
\caption{\textcolor{black}{\virg{Effective} general relativistic  parts  $\Delta J^{(\rm eff)}_{\ell}$  of the odd zonals of degree $\ell=3,5,7$ obtained from the range-rate perturbations $\Delta\dot\rho$. They are another preliminary measure of a possible a-priori \virg{imprinting} of GTR itself, not explicitly solved-for  in all the GRACE-based solutions produced so far, on the odd zonals. Compare these figures with the formal, statistical errors $\sigma_{\overline{C}_{\ell,0}},\ \ell=3,5,7$ in Table \ref{zona}.}
}\label{imprin2}
\centering
\bigskip
\begin{tabular}{llll}
\hline\noalign{\smallskip}
Dynamical effect ($\Delta\dot\rho$) & $\Delta J^{(\rm eff)}_3$  & $\Delta J_5^{(\rm eff)}$ &  $\Delta J_7^{(\rm eff)}$  \\
\noalign{\smallskip}\hline\noalign{\smallskip}
Schwarzschild & $2\times 10^{-10}$ & $1\times 10^{-10}$ & $1\times 10^{-9}$ \\
Lense-Thirring & $2\times 10^{-13}$ & $2\times 10^{-13}$ & $1\times 10^{-12}$\\
\noalign{\smallskip}\hline\noalign{\smallskip}
\end{tabular}
\end{table*}
}

\textcolor{black}{Also in this case, the present-day level of accuracy in determining the odd zonal coefficients of the geopotential does not allow, in principle, to neglect
GTR as a potential source of a-priori aliasing.}

\textcolor{black}{As far as the LAGEOS-based tests of the Lense-Thirring effect are concerned, the issue of the a-priori \virg{imprinting} of GTR on the odd zonals is not a concern since the node of a satellite is not secularly affected by them.}
\textcolor{black}{\section{The impact of the time delay on the GRACE SST range}\lb{delay}
In this Section we treat the range shifts $\Delta\rho$ coming from the GTR-induced time delays $\Delta t$ affecting the intersatellite range.
\subsection{The gravitomagnetic time delay}\lb{delayLT}
Here we deal in detail with the effect of the gravitomagnetic field of the Earth on the propagation of the electromagnetic waves between both GRACE A and GRACE B, which is not modeled in the softwares used to process GRACE data.}

\textcolor{black}{The Lense-Thirring time delay $\Delta t_{\rm LT}$ between A and B can conveniently be written as \citep{ciufo03}
\eqi\Delta t_{\rm LT} = -\rp{2GS}{c^4}\left(\rp{1}{r_{\rm A}}+\rp{1}{r_{\rm B}}\right)\rp{\bds{\hat{S}}\bds\cdot\left(\bds{\hat{r}}_{\rm A}\bds\times\bds{\hat{r}}_{\rm B}\right)  }{1 + \bds{\hat{r}}_{\rm A}\bds\cdot\bds{\hat{r}}_{\rm B}}.\lb{ltdelay}\eqf
Generally speaking, for a pair of satellites orbiting at the altitude of GRACE it is
\eqi \rp{2GS}{c^4}\left(\rp{1}{r_{\rm A}}+\rp{1}{r_{\rm B}}\right) \sim 2.8\times 10^{-17}\ {\rm s}, \eqf corresponding to a range shift of just $8.5$ nm. Moreover, in the particular case of GRACE, the geometric factor entering \rfr{ltdelay} is quite small because both $\bds r_{\rm A}$ and  $\bds r_{\rm B}$ almost lie in a plane containing $\bds S$ as well. Indeed, it can be showed that
\eqi \rp{\bds{\hat{S}}\bds\cdot\left(\bds{\hat{r}}_{\rm A}\bds\times\bds{\hat{r}}_{\rm B}\right)  }{1 + \bds{\hat{r}}_{\rm A}\bds\cdot\bds{\hat{r}}_{\rm B}}\sim 10^{-3}-10^{-4}.\eqf
}

\textcolor{black}{
Thus, we conclude that the part of $\Delta \rho$ due to the Lense-Thirring time delay is completely negligible for the GRACE mission.
}

\textcolor{black}{\subsection{The Shapiro time delay}\lb{delayGE}
Concerning the impact of  the gravitoelectric part of the terrestrial gravitational field on the propagation of the electromagnetic waves linking GRACE A and GRACE B,  and included in the GRACE models, the Shapiro time delay \citep{shapiro64} affecting the SST range can be valuated as\footnote{See also \citep{moyer03}.}
\citep[p.164]{IERS}
\eqi\Delta t_{\rm GE}=\rp{2GM}{c^3}\ln\left(\rp{r_{\rm A} + r_{\rm B} + \rho}{r_{\rm A} + r_{\rm B} - \rho}\right).\lb{shapiro}\eqf
It can be shown that \rfr{shapiro} corresponds to a secularly increasing range shift $c\Delta t_{\rm GE}\sim 200\ {\rm \mu m}$ over $\Delta P= 1$ d.
Thus, although the Shapiro delay on the SSR range is detectable, it is about 30 times smaller than the range shift due to the orbital motions.
}
\section{Summary and conclusions}\lb{conclu}
Given the present-day high level of accuracy in measuring the GRACE satellite-to-satellite biased range ($\sigma_{\rho}\lesssim 10\ \mu$m) and range-rate ($\sigma_{\dot\rho}\lesssim 1\ \mu$m s$^{-1}$), we preliminarily investigated the \textcolor{black}{impact of GTR} on such directly observable quantities. We did not consider the general relativistic effects connected with the propagation of the electromagnetic waves linking the two spacecrafts. A further motivation for the present sensitivity analysis is given  by the currently ongoing efforts to design a follow-on of GRACE accurate to nm s$^{-1}$, or better, in the range-rate. Moreover,  GTR has never been solved-for in all the GRACE-based Earth's global gravity field solutions produced so far, so that the multipoles of the terrestrial gravitational field estimated in them may, in principle, retain an a-priori \virg{imprinting} of GTR itself. This fact is important since some of such Earth's global gravity \textcolor{black}{models} were used as background reference models in some tests of GTR itself performed with other satellites.

By numerically integrating the GRACE A/B \textcolor{black}{post-Newtonian} equations of motion in a geocentric frame over a time span 1 d long, corresponding to about 15 full orbits, we found that the GTR range signals are as large as 6000 $\mu$m (Schwarzschild) and 80 $\mu$m (Lense-Thirring), while the sizes of the range-rate GTR effects are $10$ $\mu$m s$^{-1}$ (Schwarzschild) and $0.012$ $\mu$m s$^{-1}$ (Lense-Thirring). \textcolor{black}{An analytical calculation for the Lense-Thirring effect confirmed the numerical results concerning it.}
If, on the one hand, they are larger than $\sigma_{\rho}$ and $\sigma_{\dot\rho}$, apart from the Lense-Thirring range-rate, on the other hand the imperfect knowledge of some low-degree zonal harmonics of the geopotential causes competing range and range-rate perturbations which would corrupt the recovery of the relativistic signals of interest. According \textcolor{black}{to} the formal, statistical errors released in one of the latest GOCE/GRACE-based models, the mismodelled signatures of the first six zonals are smaller than the Schwarzschild range and range-rate ones by several orders of magnitude. Instead, they are about of the same order of magnitude of, or slightly smaller than, the Lense-Thirring range and range-rate perturbations.
Conversely, by comparing the relativistic and the zonal orbital effects on the GRACE range and range-rate it was possible to quantitatively assess the level of a possible a-priori \virg{imprinting} of GTR itself on the zonals considered. It turned out that it is not negligible as far as the Schwarzschild component is \textcolor{black}{concerned},
while the Lense-Thirring  \virg{imprint}  is about at the edge of the present-day level of accuracy in determining them.

As a caveat concerning the actual measurability of the GTR effects considered here, we stress the need of checking with extensive numerical simulations if the relativistic signatures are not  absorbed and removed from the range signal in estimating some of the various range parameters which are solved-for in the usual GRACE data processing. Implementing such a non-trivial task is outside the scopes of the present paper, but it could-and, in fact, should-be done by the various groups worldwide engaged in analyzing longer and longer GRACE data records. It is highly desirable that they produce dedicated global gravity field models  explicitly estimating GTR as well along with the usual Stokes coefficients of the geopotential.

Finally, let us note that the approach presented here can, in principle, also be extended to other satellite-to-satellite orbital configurations suitably designed to enhance the relativistic signatures, and to the dynamical effects caused by various modified models of gravity. \textcolor{black}{Such a possibility has recently become appealing after the proposal  of a putative Earth-sourced fifth force as a possible explanation of Earth-based neutrino phenomenology}.

\end{document}